\input harvmac
%
%
\newcount\figno
\figno=0
\def\fig#1#2#3{
\par\begingroup\parindent=0pt\leftskip=1cm\rightskip=1cm\parindent=0pt
\baselineskip=11pt \global\advance\figno by 1 \midinsert
\epsfxsize=#3 \centerline{\epsfbox{#2}} \vskip 12pt {\bf Fig.\
\the\figno: } #1\par
\endinsert\endgroup\par
}
\def\figlabel#1{\xdef#1{\the\figno}}
\overfullrule=0pt
\parskip=0pt plus 1pt
\sequentialequations

\def\ZZ{\hbox{\bf Z}}

\def\Qp{\hbox{\bf Q}_p}

\def\dnd{|\!\!/}
\def\np{Nucl.\ Phys.}
\def\pl{Phys.\ Lett.}

\def\mpl{Mod.\ Phys.\ Lett.}

\def\jhep{JHEP} 
\Title{\vbox{\baselineskip12pt\hbox{math-ph/0606003}%
\hbox{HRI-P-0604001}%
\hbox{AEI-2006-036}
}}%
{\vbox{\centerline{Quantum Extended Arithmetic Veneziano Amplitude}}}

{\vskip -20pt\baselineskip 14pt

\centerline{Debashis Ghoshal} 

\bigskip
\centerline{\it Harish-Chandra Research Institute}
\centerline{\it Chhatnag Road, Jhusi}
\centerline{\it Allahabad 211 019, India}
\smallskip
\centerline{\tt ghoshal@mri.ernet.in}
\smallskip

\vglue .3cm
\bigskip\bigskip

\noindent The Veneziano amplitude for the tree-level scattering of
four tachyonic scalar of open string theory has an arithmetic analogue
in terms of the $p$-adic gamma function. We propose a quantum
extension of this amplitude using the $q$-extended $p$-adic gamma
function given by Koblitz. This provides a one parameter deformation
of the arithmetic Veneziano amplitude. We also comment on the
dificulty in generalising this to higher point amplitudes.  }
\Date{May 2006}

\lref\SenNF{
A.\ Sen, {\it Tachyon dynamics in open string theory}, Int.\ J.\ Mod.\ Phys.\ 
{\bf A20} (2005) 5513, [{\tt hep-th/0410103}].}

\lref\Vene{
G.\ Veneziano, {\it Construction of a crossing-symmetric, Regge behaved amplitude 
for linearly rising trajectories}, Nuovo Cim.\ {\bf A57} (1968) 190.}

\lref\CoonYW{
D.\ Coon, {\it Uniqueness of the Veneziano representation}, \pl\ {\bf B29} (1969) 
669.}

\lref\BakerCoon{
M.~Baker and D.~D.~Coon,
{\it Dual Resonance Theory With Nonlinear Trajectories},
Phys.\ Rev.\ D {\bf 2}, 2349 (1970).
}

\lref\RomansQS{
L.\ Romans,  {\it Deforming the Veneziano model}, in Proceedings, High energy 
physics and cosmology, 278-287, Trieste (1989); also 
{\it A new family of dual models (`Q Strings')}, Preprint USC88/HEP-014 
(as quoted in proc.\ above).}

\lref\qVene{M.\ Chaichian, J.\ Gomes and P.\ Kulish,
{\it Operator formulation of q-deformed dual string model}, \pl\ {\bf B311}
(1993) 93;\hfill\break
M.\ Chaichian, J.\ Gomes and R.\ Gonzalez-Felipe,
{\it New phenomenon of nonlinear trajectory and quantum dual string theory}, \pl\ 
{\bf B341} (1994) 147.}

\lref\twodefVene{
L.\ Jenkovszky, M.\ Kibler and A.\ Mishchenko, {\it Two parametric quantum 
deformed dual amplitude}, \mpl\ {\bf A10} (1995) 51, 
[{\tt hep-th/9407071}].}

\lref\FairlieAD{
D.\ Fairlie and J.\ Nuyts, {\it A fresh look at generalized Veneziano amplitudes},
\np\ {\bf B433} (1995) 26 [{\tt hep-th/9406043}].}

\lref\noghostQstr{
N.\ Mebarki, A.\ Boudine, H.\ Aissaoui, A.\ Benslama, A.\ Bouchareb 
and A.\ Maasmi, {\it No-ghost theorem for a q-deformed open bosonic string}, J.\ 
Geom.\ Phys.\  {\bf 30} (1999) 103.}

\lref\FrOl{P.G.O.\ Freund and M.\ Olson,
{\it Nonarchimedean strings}, \pl\ {\bf B199} (1987) 186.}

\lref\FrWi{P.G.O.\ Freund and E.\ Witten, {\it Adelic string 
amplitudes}, \pl\ {\bf B199} (1987) 191.}

\lref\BFOW{L.\ Brekke, P.G.O.\ Freund, M.\ Olson and 
E.\ Witten, {\it Non-archimedean string dynamics},
Nucl.\ Phys.\  {\bf B302} (1988) 365.}

\lref\OtherPadic{P.H.\ Frampton and Y.\ Okada,
{\it The p-adic string N point function},
Phys.\ Rev.\ Lett.\  {\bf 60} (1988) 484;
{\it Effective scalar field theory of p-adic string},
Phys.\ Rev.\  {\bf D37} (1988) 3077; \hfill\break
P.H.\ Frampton, Y.\ Okada and M.R.\ Ubriaco,
{\it On adelic formulas for the p-adic string},
Phys.\ Lett.\  {\bf B213} (1988) 260.}

\lref\BrFrRev{
L.\ Brekke and P.G.O.\ Freund, {\it p-Adic numbers in physics}, 
Phys.\ Rep.\ {\bf 233} (1993) 1, and references therein.}

\lref\GGPSP{
I.M.~Gelfand, M.I.~Graev and I.I.~Pitaetskii-Shapiro,
{\it Representation theory and automorphic functions}, Saunders (1969).}

\lref\RussP{
V.S.~Vladimirov, I.V.~Volovich and E.I.~Zelenov,
{\it p-Adic Analysis and Mathematical Physics},
(Series on Soviet and East European Mathematics, Vol.1) World Scientific,
1994. }

\lref\GhSeP{D.\ Ghoshal and A.\ Sen, {\it Tachyon condensation and brane 
descent relations in $p$-adic string theory}, \np\ {\bf B584} (2000) 
300, [{\tt hep-th/0003278}].}

\lref\GhoNCP{D.\ Ghoshal, {\it Exact noncommutative solitons in $p$-adic 
strings and BSFT}, \jhep\ {\bf 0409} (2004) 41, 
[{\tt hep-th/0406259}].}

\lref\GhKaBP{D.\ Ghoshal and T.\ Kawano, {\it Towards $p$-adic string in 
constant $B$-field}, \np\ {\bf B710} (2005) 577, [{\tt hep-th/0409311}].}

\lref\FrRev{For a review including recent developments, see
P.\ Freund, {\it p-adic strings and their applications},
[{\tt hep-th/0510192}].}

\lref\GhIs{
D.\ Ghoshal, {\it p-Strings vs.\ strings}, Preprint HRI-P-0605001, to
appear in the proceedings of the `12th Regional Conference in
Mathematical Physics', Islamabad, Pakistan, World Scientific.}
 
\lref\VoloP{I.V.\ Volovich, {\it p-Adic string},
Class.\ Quant.\ Grav.\  {\bf 4} (1987) L83.}

\lref\GrossP{B.\ Grossman, {\it p-Adic strings, the Weyl conjectures and anomalies},
Phys.\ Lett.\  {\bf B197} (1987) 101.}

\lref\KobP{
N.~Koblitz,
{\it p-Adic numbers, p-adic analysis and zeta functions},
GTM 58, Springer-Verlag (1977).}


\lref\RobP{
A.M.~Robert,
{\it A course in p-adic analysis}, GTM 198, Springer-Verlag (2000).}

\lref\CandelasFQ{
P.~Candelas, X.~de la Ossa and F.~Rodriguez-Villegas,
{\it Calabi-Yau manifolds over finite fields, I}, [{\tt hep-th/0012233}];
{\it Calabi-Yau manifolds over finite fields, II}, [{\tt hep-th/0402133}].}

\lref\SchimmrigkZQ{
R.~Schimmrigk, {\it Arithmetic of Calabi-Yau varieties and rational conformal field  
theory}, J.\ Geom.\ Phys.\  {\bf 44} (2003) 555, [{\tt hep-th/0111226}];
{\it Arithmetic spacetime geometry from string theory}, [{\tt hep-th/0510091}].}

\lref\KontsevichAN{
M.~Kontsevich, A.~Schwarz and V.~Vologodsky, {\it Integrality of instanton numbers 
and p-adic B-model}, [{\tt hep-th/0603106}].}

\lref\KobQPGamma{N.\ Koblitz, {\it q-Extension of the p-adic gamma function},
Trans.\ Am.\ Math.\ Soc.\ {\bf 260} (1980) 449.}

%

{\nopagenumbers

\ftno=0
}

In the context of string theory, the Veneziano amplitude, which also
marks the birth of this subject, describes the tree level scattering
of four tachyonic scalars of the open bosonic string. The tachyon
field, localised on a D-brane defined by the open string, signals an
instability of this classical vacuum configuration. Understanding the
dynamics of this field has been one of the subject of foremost
interest in recent years and has provided us with valuable insight
into non-perturbative aspect of string theory (see \refs{\SenNF} and
references therein). 

This motivates one to consider the tachyon amplitudes from many
different perspectives. Mathematically there are several possibilities.
One that was considered soon after Veneziano's proposal\refs{\Vene} is
a quantum extension or $q$-deformation\refs{\CoonYW,\BakerCoon} using
the $q$-deformed gamma and beta functions. This was further
investigated in the following years (see \refs{\RomansQS} for history
and a list of references) and revisited more recently in
\refs{\RomansQS\qVene\twodefVene\FairlieAD--\noghostQstr}. 

Another direction of investigation considered the problem in the
domain of local fields. In fact, two quite different theories are
possible.  The first starts with the integral representation of the
Veneziano amplitude (and its generalisation the Koba-Nielsen
amplitudes) and defines their $p$-adic
analogues\refs{\FrOl\FrWi\BFOW--\OtherPadic} by appropriately
generalising the formulas to the field $\Qp$ of $p$-adic numbers (see
\refs{\BrFrRev} for a review). The Veneziano amplitude, for example,
involves complex valued Gelfand-Graev gamma function on the $p$-adic
field\refs{\GGPSP,\RussP}. In this theory, which we may call the {\it
$p$-adic string} theory, the spacetime is the usual one and the
properties of the tachyon qualitatively resembles those of the usual
bosonic string\refs{\GhSeP\GhoNCP\GhKaBP\FrRev--\GhIs}. In the second
approach\refs{\VoloP,\GrossP}, one directly defines the $p$-adic
analogues of the Veneziano formula in terms of Morita's $p$-adic
valued $p$-gamma function\refs{\KobP,\RobP}. This latter version will
be called {\it arithmetic Veneziano amplitude} to distinguish it from
the former one. Other arithmetic analogues of string theory can be
found in Refs.\refs{\CandelasFQ\SchimmrigkZQ--\KontsevichAN}.

In this note, we consider a quantum extension of the {\it arithmetic
Veneziano amplitude} using the $q$-extension of the $p$-adic gamma
function given by Koblitz\refs{\KobQPGamma}. 


\bigskip

Morita's $p$-adic gamma function 
$\Gamma_p\,:\,\ZZ_p\rightarrow\ZZ_p^*$ is defined on positive 
integers as
\eqn\moritaone{
\Gamma_p(n+1) = (-)^{n+1} \prod_{{m=1\atop p\dnd m}}^n m,}
where, $p\dnd m$ means that $m$ is not divisible by $p$. This is then
extended to $\ZZ_p$ by continuity\refs{\KobP,\RobP} and is a
generalisation of the factorial function to $p$-adic integers. In
order to extend to all of $\Qp$, one may rewrite the above as
\eqn\moritatwo{
\Gamma_p(n+1) = (-)^{n+1} \prod_{{m=1\atop p\dnd m}}^\infty 
{m\over n+m}, }
and the continue by the replacement $n\to x\in\Qp$. The $p$-adic 
gamma function satisfies the recursion relation
\eqn\mgammarecur{
\Gamma_p(x+1) = \cases{-\,x \Gamma_p(x), &if $x\in\ZZ_p^*$,\cr
-\,\Gamma_p(x), &if $x\in p\ZZ_p$.} }

A $q$-extension of the above is defined by Koblitz\KobQPGamma. One 
again starts with its form for positive integers:
\eqn\qpgammaZ{
\Gamma_{p,q}(n+1) = (-)^{n+1} \prod_{{m=1\atop p\dnd m}}^n 
{1-q^m\over 1-q}, }
where, $0<|q-1|_p<1$; and then extends to $\Qp$ by 
continuity\foot{Actually $q$ here may be an element of 
$\hbox{\bf C}_p$, the completion of the algebraic closure of $\Qp$.}:
\eqn\qpgamma{
\Gamma_{p,q}(x+1) = (-)^{x+1}(1-q)^{-x}\prod_{{m=1\atop p\dnd m}}^\infty 
{1-q^m\over 1-q^{x+m}}.}
Various properties of $\Gamma_{p,q}(x)$ are studied in \refs{\KobQPGamma}. 
For example,
\eqn\qpgammarecur{
\Gamma_{p,q}(x+1) = \cases{-\,{1-q^x\over 1-q}\, \Gamma_{p,q}(x), 
&if $x\in\ZZ_p^*$,\cr
-\,\Gamma_{p,q}(x), &if $x\in p\ZZ_p$,} }
and $\Gamma_{p,q}(1)=-1$. 

All these parallel the case of the usual gamma function and its 
$q$-analogue $\Gamma_q(x)$
\eqn\qgamma{
\Gamma_q(x+1) 
= (1-q)^{-x} \prod_{m=1}^\infty {1-q^m\over 1-q^{x+m}}.}
The functions $\Gamma_q(x)$ appear in the $q$-deformed Veneziano
amplitude\refs{\CoonYW\BakerCoon\RomansQS\qVene\twodefVene\FairlieAD--\noghostQstr}.
(This amplitude also has an infinite number of complex poles
distributed along the imaginary axis for each pole at real momentum
values.  This is reminiscent of the singularity structure of the other
kind of $p$-adic string amplitudes\refs{\FrOl\FrWi\--\BFOW}.)
It is well known that in the $q\to 1$ (classical) limit, one recovers
the undeformed function: $\lim_{\scriptstyle{q\to 1}}\,
\Gamma_q(x)\,=\,\Gamma(x)$. Through this relation one recovers the
usual Veneziano amplitude in the classical limit.

Similarly in the $p$-adic case,  
\eqn\climit{
\lim_{q\to 1} \Gamma_{p,q}(x)\,=\,\Gamma_p(x).}
Therefore, the $q$-extended function $\Gamma_{p,q}$ is a one parameter
family of deformation of the function \moritaone. Since, the only
difference between \moritaone\ or \qpgammaZ\ and their usual 
counterparts are from those factors larger than $p$,  one finds that
\eqn\pinfty{
\lim_{p\to\infty} \Gamma_p(n) = (-)^n\Gamma(n),\qquad\qquad
\lim_{p\to\infty} \Gamma_{p,q}(n) = (-)^n\Gamma_q(n), }
{\it i.e.}, a passage to the $p$-adic and usual gamma functions
evaluated at integers.


\bigskip

The arithmetic Veneziano amplitude were written in terms of the
$p$-adic gamma function \moritatwo. Consider $d$-dimensional `momenta'
$k_1,\cdots, k_4$ valued in the vector space $\Qp^d$ and a
$\Qp$-valued quadratic form $\langle .|.\rangle$ on it such that
$k_i^2\equiv\langle k_i|k_i\rangle = 2$. Let $s=(k_1+k_2)^2$ and
$t=(k_1+k_3)^2$ are the Mandelstam variables and define linear
functions $\alpha(s)=1+\alpha's$ and similarly $\alpha(t)$.  The
proposed form of the amplitude is\refs{\VoloP,\GrossP}
\eqn\pvene{
A_p(s,t) = {\Gamma_p\left(\alpha(s)\right)\Gamma_p\left(\alpha(t)\right)
\over\Gamma_p\left(\alpha(s)+\alpha(t)\right)}. }
This amplitude has poles whenever the momenta in the intermediate
channel is (proportional to) a negative integer not divisible by $p$. 
There are an infinite number of poles in either the $s$ or the $t$
channel. 

A straightforward quantum extension of the above can be written
in terms of the $q$-deformed $p$-adic gamma function \qpgamma:
\eqn\qpvene{
A_{p,q}(s,t) = {\Gamma_{p,q}\left(\alpha(s)\right)\Gamma_{p,q}\left(
\alpha(t)\right)\over\Gamma_{p,q}\left(\alpha(s)+\alpha(t)\right)}.}
{}From its definition and the property \climit, it is immediately
obvious that the above gives us a one-parameter family of arithmetic
Veneziano amplitude. Moreover, in the $p\to\infty$ limit thanks to
\pinfty, the $q$-deformed arithmetic family makes contact to the 
usual $q$-deformed Veneziano amplitude. The analytic structure of
the amplitude \qpvene\ is an expected combination of that of
the $p$-adic amplitude \pvene\ and the usual $q$-deformed one.

\bigskip

Finally, let us discuss about the difficulty in generalising the
arithmetic Veneziano amplitude and its quantum extension to the higher
point tree amplitudes. Recall that, in the usual case, one can use the
symbol $(a;q)_n=\displaystyle\prod_{m=0}^{n-1}\left(1-q^m a\right)$
(also valid in the limit $n\to\infty$), to write the $q$-gamma
function as $\Gamma_q(x) =
(1-q)^{1-x}(q;q)_\infty/(q^{x};q)_\infty$. There is a useful theorem,
called the $q$-binomial theorem,
\eqn\qbinth{
{(zq^\alpha;q)_\infty\over (z;q)_\infty} = \sum_{m=0}^\infty
{(q^\alpha;q)_m\over (q;q)_m}\, z^m, }
which may be used to re-write the $q$-deformed Veneziano amplitude
as follows.
\eqn\qBTinqVene{
\eqalign{
A_{q}(s,t) &= (1-q)(q;q)_{\infty} {\left(q^{\alpha(s)+\alpha(t)};
q\right)_{\infty}\over \left(q^{\alpha(s)};q\right)_{\infty}
\left(q^{\alpha(t)};q\right)_{\infty}}\cr 
&= (1-q)(q;q)_{\infty} 
\sum_{m} {q^{m\alpha(s)}\over (q;q)_{m}
\left(q^{m + \alpha(t)};q\right)_{\infty} }\cr
&= (1-q)(q;q)_{\infty} 
\sum_{m, n} {q^{m\alpha(s) + n\alpha(t) + mn}
\over (q;q)_{m}(q;q)_{n}}. }}
In the second line above, one sees the poles in the $t$-channel ---
alternatively one could have displayed the singularities in the
$s$-channel. More interesting is the symmetric form in the last line.
It suggests a possible generalisation to the higher point
amptitudes\refs{\BakerCoon,\RomansQS}. For the $n$-point amplitude,
consider the set of $n(n-3)/2$ independent planar channels labelled by
$\{i,j\},\;1\le i<j< n$ (excluding $\{1,n-1\}$) corresponding to the
set of tachyons numbers $i,i+1,\cdots,j$. Let,
$s_{ij}=\left(p_i+p_{i+1}+\cdots+p_j\right)^2$ and
$\alpha_{ij}\equiv\alpha\left(s_{ij}\right)$. The $n$-point tree
amplitude is:
\eqn\qnpoint{
{\cal A}^{(n)}_q = \left[(1-q)(q;q)_\infty\right]^{n-3}
\sum_{\{\ell\}}\,\prod_{ij}{q^{\ell_{ij}\alpha_{ij}}
\over (q;q)_{\ell_{ij}}}\;\prod_{mn;m'n'} q^{\ell_{mn}\ell_{m'n'}}, }
where, $\ell_{ij}$ is the summation index in the $ij$ channel and the
first product above is over all single channels and the second over
all distinct pairs of overlapping ones.

Let us try to follow the above steps as closely as possible in the 
$p$-adic case. First in writing Koblitz' $q$-extended $p$-gamma function 
\qpgamma, we are naturally led to the symbol
\eqn\psymbol{
(a;q)_{p,n}=\displaystyle\prod_{{m=0\atop p\dnd m+1}}^{n-1}
\left(1-q^m a\right), }
in terms of which
\eqn\kgammainsym{
\Gamma_{p,q}(x) = (-)^x
(1-q)^{1-x}\,{\left(q;q\right)_{p,\infty}\over
\left(q^x;q\right)_{p,\infty}}. }
Let us look for a generalisation of the $q$-binomial theorem starting
with the ratio of $p$-adic symbols:
\eqn\ratio{
{\left(q^\alpha z;q\right)_{p,\infty}\over\left(z;q\right)_{p,\infty}}
= \prod^\infty_{{m=0}\atop p\dnd m+1}{1-q^{m+\alpha}z\over 1-q^m z}.}
For an integer $\alpha$, i.e., $\alpha=n$ in the $q\to 1$ limit the
above reduces to $\left(1-z\right)^{-n+\nu_p(n)}$, where the function 
$\nu_p(n)=\displaystyle{\sum_{m\le n,p| m}}1$,
counts the numbers upto and including $n$ that are divisible by the
given prime $p$. Incidentally, $\nu_p(n)$ seems to be inherently more
arithmetic that what we have encountered so far. Using standard tricks
of manipulation of infinite products, we remove the 
restriction by putting in extra terms and compensate
for it by another product running over integers which are multiples
of $m+1$. This leads to
\eqn\pratiotoratio{
{\left(q^\alpha z;q\right)_{p,\infty}\over\left(z;q\right)_{p,\infty}}
= {\left(q^\alpha z;q\right)_\infty/\left(z;q\right)_\infty\over
\left.\left({q'}^{\alpha/p} z';q'\right)_\infty\right\slash
\left(z';q'\right)_\infty}, }
where, $q'=q^p$ and $z=q^{p-1}z$. Now one can use the $q$-binomial
theorem \qbinth\ to simplify this. In particular, the $q$-extended
$p$-adic Veneziano amplitude is
\eqn\qpvtoqv{
\eqalign{
A_{p,q}(s,t) &= (1-q)\left(q;q\right)_{p,\infty}\;
{\left(q^{\alpha(s)+\alpha(t)};q\right)_{p,\infty}\over
\left(q^{\alpha(s)};q\right)_{p,\infty}
\left(q^{\alpha(t)};q\right)_{p,\infty}}\cr
&= {(1-q)\left(q;q\right)_{p,\infty}\;
\left(q^{\alpha(s)+\alpha(t)};q\right)_{\infty}\left\slash
\left(q^{\alpha(s)};q\right)_{\infty}
\left(q^{\alpha(t)};q\right)_{\infty}\right.\over 
\left.\left({q'}^{(\alpha(s)+\alpha(t))/p}q^{p-1};q'\right)_{\infty} 
\right\slash\left({q'}^{\alpha(s)/p}q^{p-1};q'\right)_{\infty}
\left({q'}^{\alpha(t)/p}q^{p-1};q'\right)_{\infty}}, }}
which is expressed as a ratio of the ordinary $q$-symbols and thus in
turn can be simplified with \qbinth. Despite the initial impression,
the form above does not lead to something like the last line of
\qBTinqVene. The extra factors of $q^{p-1}$ spoil our effort. 
The case $q^{p-1}=1$ may appear to be promising, however, it gives
the trivial result $A_{p,q}(s,t)=1$.

\bigskip\bigskip

\noindent{\bf Acknowledgement:} We are grateful to Peter Freund and
Stefan Theisen for discussion. It is a pleasure to acknowledge the
hospitality of the Albert Einstein Institute, Potsdam, Germany, where
this work was done.


\listrefs 

\bye